# Researcher

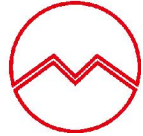

MARSLAND PRESS
Multidisciplinary Academic Journal Publisher

## Body Travel Performance Improvement of Space Vehicle Electromagnetic Suspension System using LQG and LQI Control Methods

Mustefa Jibril[1], Messay Tadese[2], Eliyas Alemayehu Tadese[3]

[1] Msc, School of Electrical & Computer Engineering, Dire Dawa Institute of Technology, Dire Dawa, Ethiopia
[2] Msc, School of Electrical & Computer Engineering, Dire Dawa Institute of Technology, Dire Dawa, Ethiopia
[3] Msc, Faculty of Electrical & Computer Engineering, Jimma Institute of Technology, Jimma, Ethiopia

**Abstract** Electromagnetic suspension system (EMS) is mostly used in the field of high-speed vehicle. In this paper, a space exploring vehicle quarter electromagnetic suspension system is modelled, designed and simulated using linear quadratic optimal control problem. Linear quadratic Gaussian and linear quadratic integral controllers are designed to improve the body travel of the vehicle using bump road profile. Comparison between the proposed controllers is done and a promising simulation result have been analyzed.

## 1. Introduction

Electromagnetic suspension (EMS) is the magnetic levitation of an object achieved by using continuously altering the power of a magnetic field produced through electromagnets the usage of a remarks loop. In maximum cases the levitation impact is commonly due to permanent magnets as they don't have any strength dissipation, with electromagnets simplest used to stabilize the impact. According to Earns haw's Theorem a paramagnetic ally magnetized body can't relaxation in solid equilibrium whilst located in any aggregate of gravitational and magneto static fields. In those forms of fields an equilibrium circumstance exists. Although static fields cannot provide stability, EMS works by way of always altering the modern-day sent to electromagnets to change the energy of the magnetic area and permits a solid levitation to occur. In EMS a feedback loop which constantly adjusts one or extra electromagnets to accurate the item's movement is used to cancel the instability. Many structures use magnetic attraction pulling upwards towards gravity for these styles of structures as this gives some inherent lateral stability, however a few use an aggregate of magnetic appeal and magnetic repulsion to push upwards. Magnetic levitation generation is important as it reduces electricity consumption, in large part obviating friction. It additionally avoids wear and has very low protection necessities.

## 2. Mathematical Modelling of the Electromagnetic Suspension System

Figure 1 shows a quarter vehicle electromagnetic suspension system and the electromagnetic tire design.

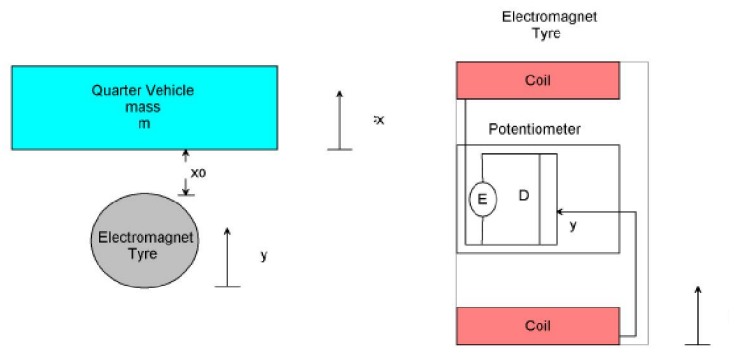

Figure 1 Quarter vehicle electromagnetic suspension system and the electromagnetic tire design.

The tire is suspended with an initial length x0 for an initial current i0. The electromagnet tire has a potentiometer fixed inside it and at the layer of the tire there is a coil connected. The potentiometer has a source voltage E and a resistor length D. When a road disturbance is accrued, the tire with the attached potentiometer rod will move upward and downward with a length y so the metallic body mass will move. The design is as follows.

Apply Kirchhoff's voltage equation for the electric circuit

$$V = V_R + V_L \Rightarrow u(t) = iR + L\frac{di}{dt} \qquad (1)$$

Where u, I, R and L is applied voltage input, current in the electromagnet coil, coil's resistance and coil's inductance respectively.

Energy stored in the inductor can be written as

$$W_{I\,Stored} = \frac{1}{2}Li^2 \qquad (2)$$

Since power in electrical system ($Pe$) = Power in the mechanical system ($Pm$),

Where $P_e = \frac{dW_{I\,Stored}}{dt}$ and $P_m = -f_m\frac{dx}{dt}$ therefore

$$f_m = -\frac{dW_{I\,Store}}{dt}\frac{dt}{dx} = -\frac{dW_{I\,Store}}{dx} \qquad (3)$$

Where $fm$ is known as electromagnet force Now substituting (2) in the equation (3),

$$\left.\begin{aligned}
f_m &= -\frac{d}{dx}\left(\frac{1}{2}Li^2\right)\\
&= -\frac{1}{2}i^2\frac{d}{dx}(L)
\end{aligned}\right\} \qquad (4)$$

Since the inductance $L$ is a nonlinear function of body travel position (x) we shall neglect the leakage flux and eddy current effects (for simplicity), so that the inductance varies with the inverse of body travel position as follows:

$$L = \frac{K}{x} \quad Where\,in\,K = \frac{\mu_0 N^2 A}{2} \qquad (5)$$

Where, $\mu_0$ is the inductance constant, A is the pole area, N is the number of coil turns and $k$ is electromagnet force constant.

$$\left.\begin{aligned}
f_m &= -\frac{1}{2}i^2\frac{d}{dx}\left(\frac{k}{x}\right)\\
&= -\frac{1}{2}i^2\left(-\frac{k}{x^2}\right)\\
\therefore f_m &= \frac{K}{2}\left(\frac{i^2}{x^2}\right)
\end{aligned}\right\} \qquad (6)$$

If $fm$ is electromagnetic force produced by input current, $fg$ is the force due to gravity and $f$ is net force acting on the vehicle body, the equation of force can be written as

$$\left.\begin{aligned}
f_g &= f_m + f\\
&= f_m + m\left(\frac{d^2x}{dt^2}\right)\\
\Rightarrow m\frac{dv}{dt} &= f_g - f_m = mg - \frac{K}{2}\left(\frac{i(t)}{x(t)}\right)^2
\end{aligned}\right\} \qquad (7)$$

Where $m$ = vehicle mass and $v$ =dx/dt =dh/dt, which is velocity of the vehicle body movement.

At equilibrium the force due to gravity and the magnetic force are equal and oppose each other so that the vehicle body levitates. i.e. $fg = -fm$ and $f$=0. On the basis of electro-mechanical modeling, the nonlinear model of magnetic levitation system can be described as follows:

The general form of an affine system

$$\frac{dz}{dt} = f(z) + g(z).u \qquad (8)$$

Is obtained by denoting variables for state space representation as follows

$$\left.\begin{aligned}
z_1 &= x\\
z_2 &= \frac{dx}{dt} = v\\
z_3 &= i
\end{aligned}\right\} \qquad (9)$$

Substitute equation (9) or the state variables in to equation (1) and (7)

$$\left.\begin{aligned}
u(t) &= z_3.R + L.x\dot{z}_3\\
m\dot{z}_2 &= m.g - \frac{K}{2}\left(\frac{z_3}{z_1}\right)^2
\end{aligned}\right\} \qquad (10)$$

Then the nonlinear state space model is

$$\begin{aligned}
\dot{z}_1 &= x_2 \\
\dot{z}_2 &= \left(g - \frac{K}{2m}\right)\left(\frac{z_3}{z_1}\right)^2 \\
\dot{z}_3 &= \frac{u}{L} - z_3 \frac{R}{L}
\end{aligned}\right\} \qquad (11)$$

The $fm$ is electromagnetic force produced by input current is related to the current and the road disturbance displacement will be

$$f_m = g(x, i) \qquad (12)$$

Using Taylor series linearization technique we have

$$f_m = \left(\frac{\partial g}{\partial x}\right)_{x_0, i_0} x + \left(\frac{\partial g}{\partial i}\right)_{x_0, i_0} i = -\frac{2ki_0^2}{x_0^3} x + \frac{2ki_0}{x_0^2} i \qquad (13)$$

Where $g = g(x, i)$ and $(x_0, i_0)$ is the operating point's initial inputs.

The applied force to the mass become

$$M\frac{d^2 x}{dt^2} = f_m = -\frac{2ki_0^2}{x_0^3} x + \frac{2ki_0}{x_0^2} i \qquad (14)$$

Then the linearized state space model becomes

$$\begin{pmatrix} \dot{z}_1 \\ \dot{z}_2 \\ \dot{z}_3 \end{pmatrix} = \begin{pmatrix} 0 & 1 & 0 \\ -\dfrac{2ki_0^2}{mx_0^3} & 0 & \dfrac{2ki_0}{mx_0^2} \\ 0 & 0 & -\dfrac{R}{L} \end{pmatrix} \begin{pmatrix} z_1 \\ z_2 \\ z_3 \end{pmatrix} + \begin{pmatrix} 0 \\ 0 \\ \dfrac{1}{L} \end{pmatrix} v$$

$$y = \begin{pmatrix} 1 & 0 & 0 \end{pmatrix} \begin{pmatrix} z_1 \\ z_2 \\ z_3 \end{pmatrix}$$

From the potentiometer

$$v = E\frac{y}{D} \qquad (15)$$

So the final state space model becomes

$$\begin{pmatrix} \dot{z}_1 \\ \dot{z}_2 \\ \dot{z}_3 \end{pmatrix} = \begin{pmatrix} 0 & 1 & 0 \\ -\dfrac{2ki_0^2}{mx_0^3} & 0 & \dfrac{2ki_0}{mx_0^2} \\ 0 & 0 & -\dfrac{R}{L} \end{pmatrix} \begin{pmatrix} z_1 \\ z_2 \\ z_3 \end{pmatrix} + \begin{pmatrix} 0 \\ 0 \\ \dfrac{E}{LD} \end{pmatrix} y$$

$$y = \begin{pmatrix} 1 & 0 & 0 \end{pmatrix} \begin{pmatrix} z_1 \\ z_2 \\ z_3 \end{pmatrix}$$

The system parameters are shown in Table 1 below

Table 1 System parameters

| No | Parameter | Symbol | Value |
|---|---|---|---|
| 1 | Mass of the vehicle | m | 1 kg |
| 2 | Coil resistance | R | $10\ \Omega$ |
| 3 | Coil inductance | L | 0.2 H |
| 4 | Initial current | $i_0$ | 0.8 A |
| 5 | Initial displacement | $x_0$ | 0.03 m |
| 6 | Electromagnet Constant | k | $\dfrac{Nm^2}{A^2}$ 2.9x10^-6 |
| 7 | Potentiometer voltage | E | 5 V |
| 8 | Potentiometer distance | D | 0.13 m |

The transfer function become

$$\frac{X(s)}{Y(s)} = \frac{1}{s^3 + 50s^2 + 0.1375s + 6.874}$$

The state space representation becomes

$$\dot{x} = \begin{pmatrix} -50 & -0.1375 & -6.8740 \\ 1 & 0 & 0 \\ 0 & 1 & 0 \end{pmatrix} x + \begin{pmatrix} 1 \\ 0 \\ 0 \end{pmatrix} u$$

$$y = \begin{pmatrix} 0 & 0 & 1 \end{pmatrix} x$$

## 3. Proposed Controllers Design
### 3.1 LQG optimal controller Design

LQG computes an optimal controller to stabilize the plant G (s)

$$\dot{x} = Ax + Bu + \xi \qquad (16)$$
$$y = Cx + Du + \theta$$

And minimize the quadratic cost function

$$J_{LQG} = \lim_{T \to \infty} E\left\{ \int_0^T \begin{bmatrix} x^T u^T \end{bmatrix} \begin{bmatrix} Q & N_C \\ N_C^T & R \end{bmatrix} \begin{bmatrix} x \\ u \end{bmatrix} dt \right\} \qquad (17)$$

The block diagram of a quarter vehicle electromagnetic suspension system with LQG controller is shown in Figure 2 below

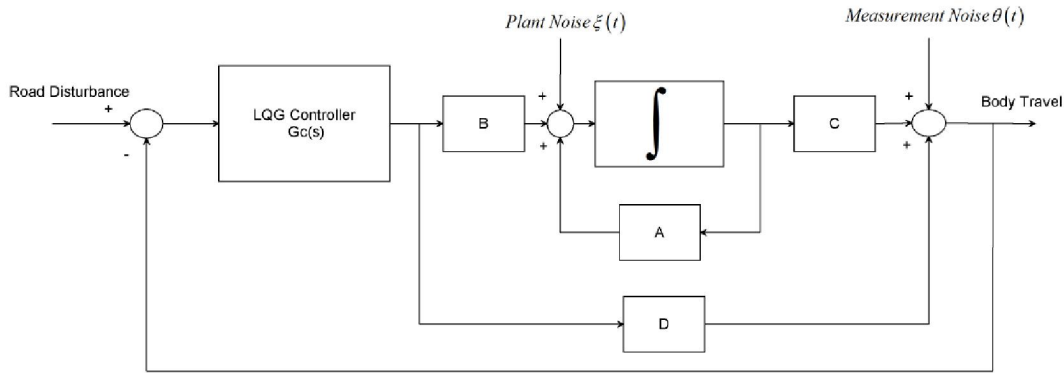

Figure 2: Block diagram of a quarter vehicle electromagnetic suspension system with LQG controller

The solution of the LQG problem is a combination of the solutions of Kalman filtering and full state feedback problems based on the so-called separation principle.

The plant noise and measurement noise are white and Gaussian with joint correlation function

$$E\left\{ \begin{bmatrix} \xi(t) \\ \theta(\tau) \end{bmatrix} \begin{bmatrix} \xi(t)\theta(\tau) \end{bmatrix}^T \right\} = \begin{bmatrix} \Xi & N_f \\ N_f^T & \Theta \end{bmatrix} \delta(t-\tau) \qquad (18)$$

The input variables W and V are

$$W = \begin{bmatrix} Q & N_C \\ N_C^T & R \end{bmatrix}; V = \begin{bmatrix} \Xi & N_f \\ N_f^T & \Theta \end{bmatrix}$$

The final negative-feedback controller becomes:

$$F(s) := \begin{bmatrix} A - K_f C_2 - B_2 K_C + K_f D_{22} K_C & K_f \\ K_C & 0 \end{bmatrix}$$

For the Gaussian noises $\Xi$ and $\Theta$ :

$$\Xi = 0.0005 \ and \ \Theta = 0.0000001$$

The value of Q and R is chosen as

$$Q = \begin{pmatrix} 5 & 0 & 0 \\ 0 & 5 & 0 \\ 0 & 0 & 5 \end{pmatrix} \ and \ R = 10$$

The LQG controller becomes

$$G_C(s) = \frac{0.5678s^3 + 2.8754s^2 + 1.4322s + 2.0123}{s^4 + 12s^3 + 35s^2 + 94s + 112}$$

### 3.2 Linear Quadratic Integral Controller Design

LQI computes an optimal state-feedback control law for the tracking loop. Block diagram of a quarter vehicle electromagnetic suspension system with LQI controller is shown in Figure 3 below.

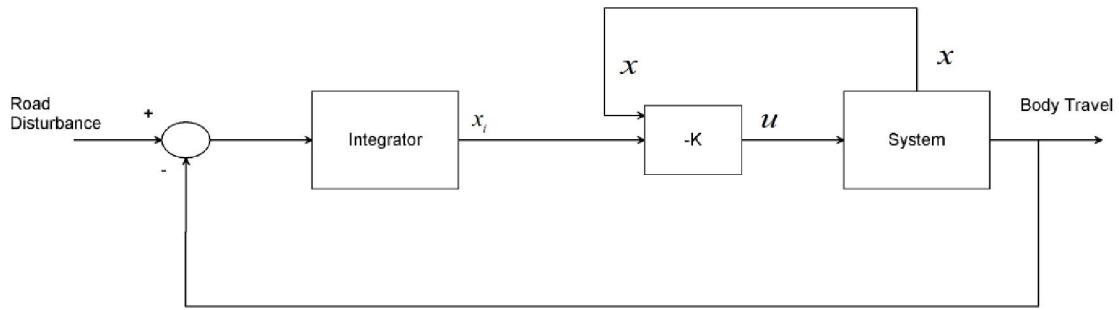

**Figure 3** Block diagram of a quarter vehicle electromagnetic suspension system with LQI controller

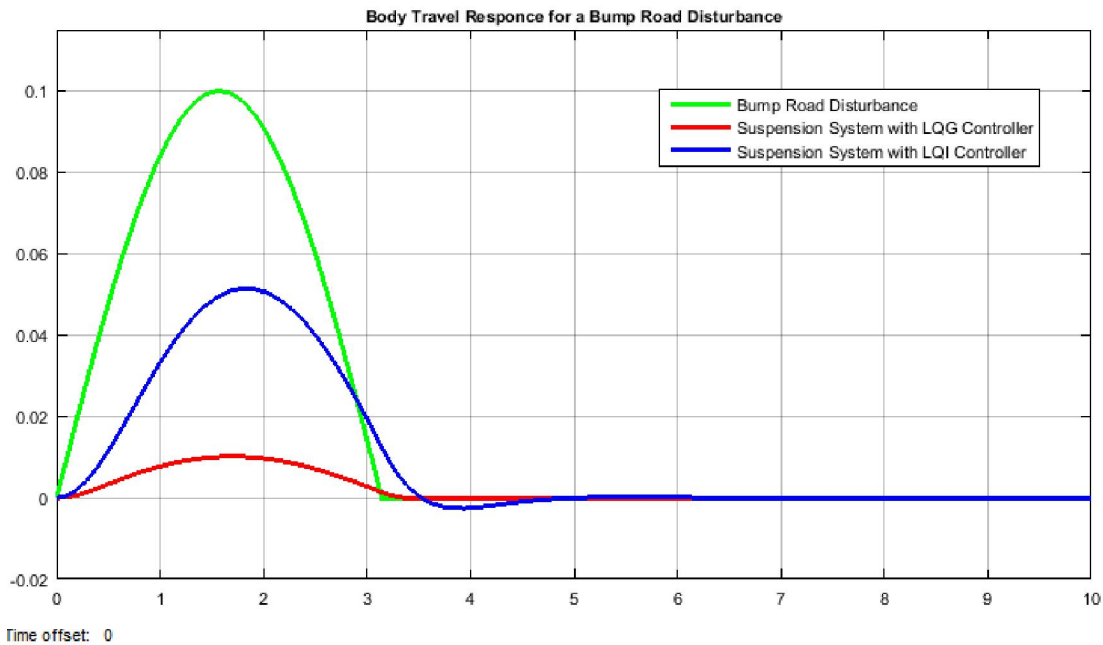

Figure 4 Body travel response for a bump road disturbance

For a plant sys with the state-space equations

$$\dot{x} = Ax + Bu \qquad (19)$$

$$y = Cx + Du$$

The state-feedback control is of the form

$$u = -K\left[x, x_i\right] \qquad (20)$$

Where $x_i$ is the integrator output. This control law ensures that the output $y$ tracks the reference command $r$. For MIMO systems, the number of integrators equals the dimension of the output $y$. LQI calculates the optimal gain matrix K, given a state-space model SYS for the plant and weighting matrices Q, R, N.

The value of Q, R and N is chosen as

$$Q = \begin{pmatrix} 5 & 0 & 0 & 0 \\ 0 & 5 & 0 & 0 \\ 0 & 0 & 5 & 0 \\ 0 & 0 & 0 & 5 \end{pmatrix} \; ; \; R = 10 \;\; and \;\; N = \begin{bmatrix} 0 \\ 1 \\ 1 \\ 1 \end{bmatrix}$$

The LQI optimal gain matrix becomes

$$K = \begin{bmatrix} 0.2004 & 9.8905 & 1.0844 & -0.7071 \end{bmatrix}$$

## 4. Result and Discussion
### 4.1 Body Travel Output Specification

One of the major specification of a suspension system is whatever the road disturbance input, the best design performance of the body travel vertical displacement is to approach to zero.

## 4.2 Comparison of a Quarter Vehicle Electromagnetic Suspension System with LQG and LQI Controllers for a Bump Road Disturbance.

The quarter vehicle electromagnetic suspension system with the proposed controller's comparison for a 10 cm bump road disturbance input simulation result for body travel response is shown in Figure 4.

The simulation result numerical value is shown in Table 1 below.

Table 2: Numerical values of the body travel simulation output

| No | Systems | Bump |
|----|--------------|--------|
| 1 | Road Profile | 0.1 m |
| 2 | LQG | 0.01 m |
| 3 | LQI | 0.05 m |

Table 2 shows that the quarter vehicle electromagnetic suspension system with LQG controller body travel is minimum and improved the road handling criteria.

## 5.    Conclusion

In this paper, a quarter vehicle electromagnetic suspension system design and analysis have been done using Matlab/Simulink. In order to increase the performance of the electromagnetic suspension system, LQG and LQI control technique is used. The main aim of this paper is to control the vehicle body travel based on road disturbance input. Comparison of the quarter vehicle electromagnetic suspension system with LQG and LQI controllers to improve the performance of the body travel output using a bump road profile. The quarter vehicle electromagnetic suspension system with LQG controller body travel is minimum and improved the road handling criteria.

**Reference**

1. Agrawal P. et al. "Design and Development of a Semi Active Electromagnetic Suspension System" SAE Technical Paper 2019-28-0110, 2019.

2. Arya, D. et al. "A Novel Approach on Electromagnetic Suspension System" International Journal of Engineering Sciences & Research Technology, Vol. 7, Issue 4, pp. 586-591, 2018.

3. Yousef A. et al. "Development of a New Automotive Active Suspension System" IOP Conference Series: Materials Science and Engineering, Vol. 280, pp. 13-15, 2017.

4. Ng J. et al. "Electromagnetic Suspension System for Control of Limb Volume in Prosthetics" Procedia CIRP, Vol. 65, pp. 180-183, 2017.

5. Sagar S. Khatavkar et al. "Electromagnetic Suspension System-A Review" International Journal of Innovative Science, Engineering & Technology, Vol. 3 Issue 3, 2016.

6. Saban Cetin "Adaptive Vibration Control of a Nonlinear Quarter Car Model with an Electromagnetic Active Suspension" Journal of Vibroengineering, Vol. 17, Issue 6, pp. 3063-3078, 2015.

7. A. I. Sultoni et al. "H ∞ Multi Objective Implementation for Energy Control of Electromagnetic Suspension" International Review of Mechanical Engineering, Vol. 9, No 6, 2015.

5/24/2020